%% file: main.tex
\documentclass[pdflatex,sn-mathphys-num]{sn-jnl}

\usepackage{graphicx}%
\usepackage{multirow}%
\usepackage{amsmath,amssymb,amsfonts}%
\usepackage{amsthm}%
\usepackage{mathrsfs}%
\usepackage[title]{appendix}%
\usepackage{xcolor}%
\usepackage{textcomp}%
\usepackage{manyfoot}%
\usepackage{booktabs}%
\usepackage{algorithm}%
\usepackage{algorithmicx}%
\usepackage{algpseudocode}%
\usepackage{listings}%
\usepackage{crypto}
\usepackage{braket}
\usepackage{bm}
\usepackage{tikz}
\usepackage{pgfplots}

\theoremstyle{thmstyleone}%
\theoremstyle{thmstyletwo}%

\theoremstyle{thmstylethree}%

\newcommand{\ignore}[1]{}

\begin{document}

\title{
Experimental Secure Multiparty Computation from Quantum Oblivious Transfer with Bit Commitment}
 

\author[1,2]{Kai-Yi Zhang}
\author[3]{An-Jing Huang}
\author[4]{Kun Tu}
\author[3]{Ming-Han Li}
\author[3]{Chi Zhang}
\author[3]{Wei Qi}
\author[1]{Ya-Dong Wu*}
\author[1,2]{Yu Yu*}
\email{wuyadong301@sjtu.edu.cn, yyuu@sjtu.edu.cn}

\affil[1]{School of Electronic Information and Electrical Engineering, Shanghai Jiao Tong University, Shanghai 200240, People’s Republic of China}
\affil[2]{State Key Laboratory of Cryptology, P.O. Box 5159, Beijing, 100878, China}
\affil[3]{Anhui CAS Quantum Network Co., Ltd., Hefei 230088, People’s Republic of China}
\affil[4]{Chengfang Financial Technology Company Limited, Beijing 100088, People’s Republic of China}


 \abstract{

Secure multiparty computation enables collaborative computations across multiple users while preserving individual privacy, which has a wide range of applications in finance, machine learning and healthcare.  Secure multiparty computation can be realized using oblivious transfer as a primitive function. 
In this paper, we present an experimental implementation of a quantum-secure quantum oblivious transfer (QOT) protocol using an adapted quantum key distribution system combined with a bit commitment scheme, surpassing previous approaches only secure in the noisy storage model. We demonstrate the first practical application of the QOT protocol by solving the private set intersection, a prime example of secure multiparty computation, where two parties aim to find common elements in their datasets without revealing any other information.
In our experiments, two banks can identify common suspicious accounts without disclosing any other data. This not only proves the experimental functionality of QOT, but also showcases its real-world commercial applications.
 }
\ignore{
 We present an experimental quantum-secure multiparty computation (MPC) protocol combining quantum oblivious transfer (QOT) and cryptographic bit commitment (BC). Building on Yao’s Millionaires’ Problem, our protocol enables parties to compute a function while maintaining privacy. We focus on cryptographic BC, choosing it over other models like noisy-storage, to ensure security in our quantum-secure MPC setup. Our experimental results demonstrate the feasibility quantum-secure multiparty computation, which may introduce new application directions to the quantum information industry.
 }

 \ignore{

What is MPC? What is OT? How to achieve 2PC from OT?

What is commitment? 

Quantum No-go Theorem.

How to break Quantum no-go theorem.

Why use commitment? Why not noisy storage model.

Crypto assumption hierarchy

Preliminary, Symbols.

[Physics part]

Performance

Appendix:

GC detail.

Parameter Detail.

Concrete failure probability analysis.

Optimize: OT Extension

Attacks : PNS attack
 
 }

\maketitle
\section{Introduction}


Secure multiparty computation (MPC) in cryptography~\cite{FOCS:Yao82b,FOCS:Yao86,STOC:GolMicWig87} enables multiple parties to jointly compute a specific function on their respective data without revealing any additional information about their private data. MPC plays a crucial role in fintech.  For example, in fraud detection and anti-money laundering, financial institutions can use MPC to jointly compute the intersection of data sets, such as blacklists and suspicious transactions, while keeping each party’s private information confidential. Beyond fintech, MPC also finds applications in privacy-preserving machine learning~\cite{SP:MohZha17}, where organizations collaborate on training models using sensitive data without exposing individual datasets. Similarly, in the secure computation of genetic data~\cite{cho2018secure}, multiple parties can perform analyses on shared genetic datasets while maintaining the privacy of participants' personal information. These use cases demonstrate MPC's versatility in facilitating secure, privacy-conscious computations across various industries.

\begin{figure*}[!ht]
    \centering
    \includegraphics[width=0.7\linewidth]{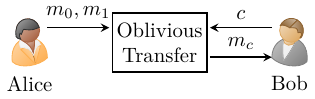}
    \caption{1-out-of-2 Oblivious Transfer. Alice has two possible messages, $m_0$	and $m_1$, to send to Bob. Bob chooses a bit 
$c\in \{0, 1\}$, and at the end of the process, Bob receives the message $m_c$, without Alice learning the value of $c$.}
    \label{fig:otexample}
\end{figure*}

Secure multiparty computation can be realized using oblivious transfer (OT) as an underlying protocol~\cite{STOC:Kilian88}. OT allows a sender to transmits data such that the receiver learns only the selected piece while the sender remains unaware of the choice. 1-out-of-2 OT, the simplest form, is illustrated in Fig.~\ref{fig:otexample}. The security of a classical OT protocol is based on the conjectured computational hardness assumptions~\cite{C:EveGolLem82,SODA:NaoPin01}, such as the RSA problem and discrete logarithm, which are vulnerable to quantum attacks using Shor's algorithm. Inspired by quantum cryptography, especially quantum key distribution (QKD), it is natural to explore a quantum analog to oblivious transfer, which we refer to as quantum OT (QOT)~\cite{santos2022quantum}. 

While unconditional security is unattainable with quantum  alone~\cite{May97,Lo97}, QOT protocols generally depend on weaker cryptographic assumptions for their security proofs compared to their classical counterparts (see Fig.~\ref{fig:hierarchy}). After QOT protocol~\cite{BBCS92} was proposed, the QOT protocol in the noisy storage model was extensively studied~\cite{PhysRevLett.100.220502,PhysRevA.81.052336,PhysRevA.82.032308} and its experimental implementation was achieved in~\cite{erven2014}. However, this protocol becomes vulnerable when large and reliable quantum memory becomes available~\cite{zhu2022demand}.
Rather than restricting the adversary's technological capability, QOT can rely on the existence of cryptographic functionality primitives such as bit commitment~\cite{JC:Naor91} to achieve security against quantum attacks.
In addition, although there were several experimental demonstrations of QOT~\cite{erven2014,furrer2018,PRXQuantum2021,PRXQuantum2023}, 
to our best knowledge, no implementations of secure multiparty computation based on QOT for addressing real-world problems have been demonstrated.

In this paper,  we experimentally implement a QOT protocol using a bit commitment scheme to ensure its theoretical security against quantum attacks.
This experimental implementation is adapted based on decoy-state QKD protocol. We then leverage our QOT protocol for multiparty secure computations.  We demonstrate its application to the private set intersection (PSI)~\cite{NDSS:HuaEvaKat12,CCS:DonCheWen13} problem, allowing two banks to securely determine the intersection of the blacklist of one bank and the new client list of the other, without revealing any additional information. 


\section{Results}
\subsection{QOT Protocol}
 
In this paper, we experimentally implement a $1$-out-of-$2$ QOT protocol and then apply it to solve one of the most practically motivated MPC problems -- the PSI problem.
Different from previous QOT protocols~\cite{PhysRevLett.100.220502,PhysRevA.81.052336,PhysRevA.82.032308,erven2014,PhysRevResearch.5.033163}, we do not assume the adversary has access only to either noisy or bounded quantum memory. Instead, our protocol is secure against any classical or quantum attacks  with polynomial time complexity.
The procedure of this protocol is summarized in Fig~\ref{fig:qot} and the detailed procedure adopting decoy-state BB84 protocol is presented in Methods~\ref{sec:protocol}. The key difference between our QOT protocol and the one in the noisy storage model is that Bob must commit all his measurement bases through a bit commitment scheme.

\begin{figure*}[!ht]
	\begin{nfprot}{QOT}
	
\begin{enumerate}

\item \textit{Alice} randomly selects $\bm{x} \in \bool^n$ and $\bm{\theta} \in \bool^n$, and prepares and sends polarization-encoded single-photon state $\ket{\bm{x}}_{\bm{\theta}}$.
\item \textit{Bob} randomly selects $\tilde{\bm{\theta}} \in \bool^n$ and measures $\ket{\bm{x}}_{\bm{\theta}}$ in the basis of $\tilde{\bm{\theta}}$, obtaining $\tilde{\bm{x}} \in \bool^n$, where \(\ket{0}_0 := \ket{0}\), \(\ket{1}_0 := \ket{1}\), \(\ket{0}_1 := \ket{+}\), and \(\ket{1}_1 := \ket{-}\). 
\item \textit{Bob} commits to all $\tilde{\bm{\theta}}, \tilde{\bm{x}}$ by sending $c_i = h(\tilde{\theta}_i, \tilde{x}_i, \bm{r}_i)$ for $i \in \{1, \dots, n\}$, where $h$ is a hash function and each $\bm{r}_i$ is a random bit string.
\item \textit{Alice} randomly selects a subset a subset \(T \subset [n]:=\{1, 2, \dots, n\}\) of size $|T|=\alpha n$, where $0<\alpha<1$. \textit{Bob} opens $c_i$ for $i \in T$, and \textit{Alice} verifies if for $i \in T$, $x_i = \tilde{x}_i$ when $\theta_i = \tilde{\theta}_i$ holds true with a ratio of at least $\beta>1/2$. If the test is passed, then they continue; otherwise, Alice aborts the protocol.
\item Both parties discard the test sets.
\item \textit{Alice} sends $\bm{\theta}$ to \textit{Bob}.
\item \textit{Bob} chooses a choice bit $c \in \bool$, divides the set $[n] \setminus T$ into good and bad ones, i.e., $I_c = \{i \mid \theta_i = \tilde{\theta}_i\}$ and $I_{1-c} = \{i \mid \theta_i \neq \tilde{\theta}_i\}$, and sends $I_0, I_1$ to \textit{Alice}. The bit strings $\bm x$ is then divided into two $\bm x_{I_c}$ and $\bm x_{I_{1-c}}$ on $I_c$ and $I_{1-c}$ respectively. Similarly, $\tilde{\bm x}$ is divided into  $\tilde{\bm x}_{I_c}$ and $\tilde{\bm x}_{I_{1-c}}$.
\item \textit{Alice} sends $\bm{y}_0 = {\sf Encode}(\tilde{\bm r}_0) \oplus (\bm{x}_{I_0})$, $\bm{z}_0 = \bm{m}_0 \oplus f_0(\tilde{\bm r}_0)$, $\bm{y}_1 = {\sf Encode}(\tilde{\bm r}_1) \oplus (\bm{x}_{I_1})$, and $\bm{z}_1 = \bm{m}_1 \oplus f_1(\tilde{\bm r}_1)$ to Bob, where $\tilde{\bm r}_0, \tilde{\bm r}_1$ are both random bit strings, ${\sf Encode}$ is the encoder of an error-correcting code, and $f_0, f_1$ are both hash functions.
\item \textit{Bob} calculates $\bm{m}_c = \bm{z}_c \oplus f_c({\sf Decode}(\bm{y}_c \oplus (\tilde{\bm{x}}_{I_c})))$, where $\sf Decode$ is the decoder of the error-correcting code.
\end{enumerate}
\end{nfprot}
\vspace{10pt}
\caption{Quantum Oblivious Transfer Protocol}
\label{fig:qot}
\end{figure*}


Here, we analyze the correctness and security of our QOT protocol, taking into account the practical limitations of quantum devices.  
The correctness of the protocol refers to the fact that both honest parties can successfully accomplish this protocol with high probability when employing the following experimental setup in Sec.\ref{sec:setup}. 
One essential imperfection introduced by Alice's quantum device is the single photon source approximated by a weak coherent state, leading to photon missing at Bob's side in certain rounds. To make the ratio of Bob's photon missing for each decoy state source fall within Alice's acceptable range, Alice should choose the range as $[p_0^{B}-\epsilon, p_0^{B}+\epsilon]$, where $p_0^{B}=p_{\ge 1}^{A}-\eta-p_{dark}$ is the expected ratio of photon clicks at Bob's detector, $p_{\ge 1}^A$ denotes the probability of generating at least one photon at Alice's decoy state source, $\eta$ denotes the transmission loss and $p_{dark}$ denotes the dark count rate of Bob's detector. To make the test pass with probability at least $1-\delta$, Alice should choose $\epsilon\ge \sqrt{1/N\log(1/\delta)}$.

Another important source of error in the protocol is bit flip errors induced by transmission noise and measurement errors. 
Bit flip errors affect two steps in the protocol: they may cause a false failure to pass Alice's test in steps 3-5 with probability $p_{\textrm{fpass}}$, and if bit flip errors occur in more than $t$ bits, they lead to failure of error correction in step 9, with probability $p_{\textrm{fcorrect}}$. The total failure probability is bounded above by
$p_{\textrm{fpass}}+p_{\textrm{fcorrect}}$. 

The security of the protocol ensures that a dishonest Alice learns no information about Bob's chosen bit $c$, while a dishonest Bob, when dealing with honest Alice, gains no knowledge of the message $\bm m_{1-c}$, which he did not select. Our OT protocol guarantees unconditional security against a malicious Alice. 
In other words, Alice does not get any information about Bob's chosen bit $c$.

For a malicious Bob, if an imperfect single-photon source emits two or more photons with nonzero probability, he could receive multiple photons during a transmission.  In this case, Bob could measure one photon immediately and store the others in his quantum memory, allowing him to measure them after Alice announces her measurement bases $\bm \theta$.
Consequently, Bob can correctly deduce the value of $x_i$  for both 
$i\in I_0$ and $i\in I_1$, thus obtaining information about 
$\bm m_{1-c}$, which he was not supposed to learn. To exploit this vulnerability, a malicious Bob could underreport the number of single-photon communications, discarding cases where he received only one photon without notifying Alice. This strategy increases the proportion of multiple-photon events compared to the total effective rounds. Such an attack is analogous to the photon number splitting (PNS) attack used by eavesdroppers in quantum key distribution (QKD)~\cite{brassard2000limitations,hwang2003quantum}. However, Bob’s PNS attack would modify the probability of photon clicks at his detector. For each decoy state, because Bob is unaware of Alice's probability 
$p_{\ge 1}^A$, his attack introduces deviations in the photon click statistics that he reports. These deviations would fall outside of Alice's acceptable range, 
 $[p_0^{B}-\epsilon, p_0^{B}+\epsilon]$, and can be used by Alice to detect Bob’s PNS attack.

We now analyze the probability that a malicious Bob can circumvent the bit commitment scheme and get both of Alice's messages. Bob might attempt to delay his measurement in step 2 until Alice announces the correct basis in step 6. However, he must first bypass the verifications in steps 3-5. Assuming a scenario most favorable for Bob, where the error rate $p_e=0$, we find that the cost for a malicious Bob to successfully bypass the bit commitment scheme and steal both of Alice's messages is exceedingly high, approximately 
 $2.8\times 10^{12}$. Here the cost is defined as  \( T/p \), where 
\( T \)  represents the running time of Bob's attack algorithm, and 
\( p \)  is the probability of a successful cheating attempt. Detailed calculations can be found in Methods~\ref{sec:CandSanalysis}.




\subsection{Experimental implementation of QOT}
\label{sec:setup}
\begin{figure*}[!ht]
    \centering
    \includegraphics[width=1\textwidth]{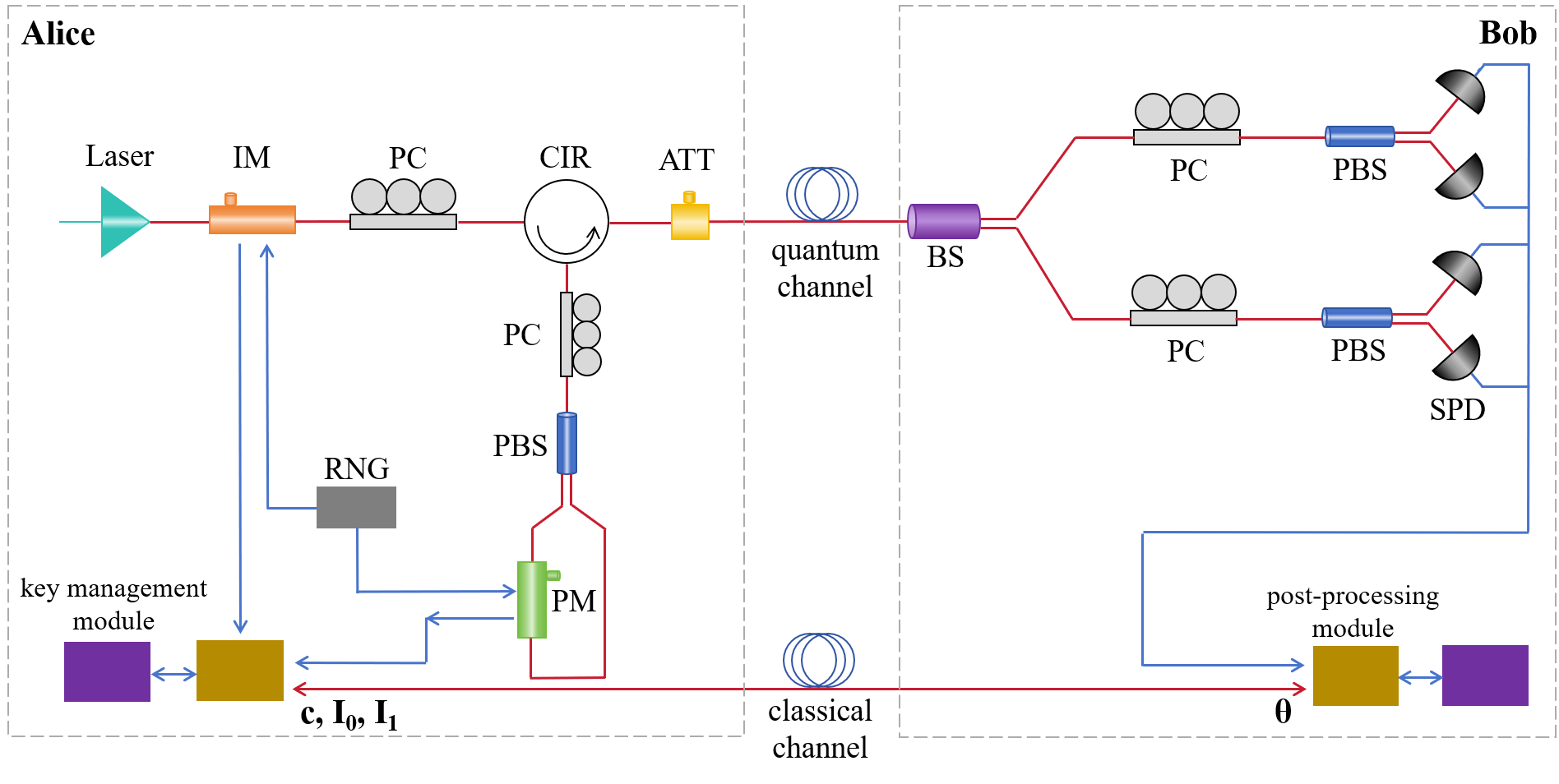}
    \caption{Schematic of experimental setup. Alice uses an intensity modulator (IM) for decoy intensity modulation. A circulator (CIR), a polarization controller (PC), a polarization beam splitter (PBS) and a phase modulator (PM) constitute the polarization encoding part. The IM and the PM are controlled by a random number generator (RNG). The light is then attenuated to a single photon level with an attenuator (ATT) and sent to Bob via fiber for detection. Bob uses a beam splitter (BS) to measure in different bases and each basis is decoded by a PC and a PBS and detected by single-photon detectors (SPDs). Post-processing and key management are performed by the corresponding modules and an extra fiber channel is used for classical communication.}
    \label{fig:setup}
\end{figure*}

We experimentally implement the above QOT protocol where both  Alice and Bob are honest. We adopt the decoy-state BB84 protocol in our experiment. Alice uses a weak coherent laser source and the decoy state method~\cite{hwang2003quantum,wang2005beating,lo2005decoy} to prevent photon number splitting attacks~\cite{brassard2000limitations,hwang2003quantum}. The schematic implementation setup is illustrated in Fig~\ref{fig:setup}. An intensity modulator (IM) is used to implement the method with three different mean photon numbers: $\mu$, $\nu$, and 0, representing signal, decoy, and vacuum states, respectively. The mean photon number of each pulse is randomly modulated among $\mu$, $\nu$, and 0 with optimized probabilities $p_1$, $p_2$, and $1-p_1-p_2$, respectively. After polarization by a polarization controller (PC), the light enters a circulator for encoding.

Polarization encoding is achieved using a PC, a polarization beam splitter (PBS), and a phase modulator (PM). The PC adjusts the angle between the input light and the optical axis of the PBS to $45^\circ$. After passing through the PBS, the two polarization components travel clockwise and counterclockwise through the same loop, are combined at the PBS, and then exit the circulator. By precisely controlling the position of the PM in the loop and the timing of the modulation signal, four BB84 polarization states are modulated. The IM and PM are both controlled by a random number generator (RNG) to ensure randomness. The signals are then attenuated to a single-photon level with an attenuator and transmitted to Bob via fiber.

On the receiving side, Bob splits the light into two paths with a beam splitter (BS) for measurement on the Z basis or X basis. Each basis is adjusted by a PC to align with the sender's basis, followed by a PBS and two single-photon detectors (SPDs) to perform projective measurements. The post-processing modules receive modulation information from the IM and PM, and detection results from the SPDs for data processing. The key management modules are used for the storage, management, input, and output of authentication keys. Two post-processing modules communicate with each other through a classical fiber channel.


We test the functionality of our QOT protocol using the above physical device. The code rates of our QOT protocol for different optical decays are shown in Table~\ref{fig:code-rate}.
\ignore{
\begin{figure}[!ht]
    \centering
\begin{tikzpicture}
  \begin{axis}[
    xlabel=Fiber Length (km),
    ylabel=Code Rate (kbps),
    xmin=0, xmax=60,
    ymin=0, ymax=80,
    axis lines=middle
  ]
    \addplot coordinates {(1,66.6) (10,66.6) (30,62.3) (50,45.7)};
  \end{axis}
\end{tikzpicture}
    \caption{Dependence of the ROT code rate and the fiber length. 
    }
    \label{fig:code-rate}
\end{figure}
}


\begin{table}[!ht]
    \begin{tabular}{@{}cccc@{}}
\toprule
Optical attenuation (dB) & Code Rate (kbps)  \\
\midrule
 -0.15 & 66.66  \\
 -1.97 & 66.11  \\
 -3.97 & 66.66  \\
 -5.96 & 66.66  \\
 -8.06 & 66.46  \\
 -9.69 & 66.66  \\
-11.81 & 50.00  \\
-13.44 & 36.66  \\
-15.58 & 25.00  \\
-17.88 & 17.77  \\
-20.21 & 11.11  \\
-22.18 & 5.00  \\
-24.42 & 3.33  \\
-26.69 & 0  \\
\bottomrule
\end{tabular}
\vspace{10pt}
\caption{Dependence of the QOT code rate on optical attenuation.}
    \label{fig:code-rate}
    \centering 
\end{table}

\subsection{Application to the private set intersection problem}
Upon completing the QOT protocol, we apply it to address a PSI problem in practice: identifying accounts blacklisted by one bank that also appear in another bank’s list of accounts, without disclosing any other clients' information.
PSI can be implemented by using an Oblivious Pseudorandom Function (OPRF) protocol, which can be further realized by our QOT protocol.

 A simplified PSI solution evolving two parties Alice and Bob using OPRF can be achieved as follows.
Denote Alice's private set as $\mathcal X$ and Bob's private set as $\mathcal Y$.
Alice first generates a secret key $\mathbf{k}$.
Alice encrypts $\mathcal X$ as $\mathcal A = \{{f(\mathbf{k},x)} : x\in \mathcal X\}$, where $f$ is a pseudorandom function, hardly distinguishable from a genuinely random function.
Then Alice and Bob execute OPRF protocol, enabling Bob, as a receiver, to evaluate $\mathcal B=\{f(\mathbf{k}, y) : y\in \mathcal Y\}$, without learning Alice's secret key $\mathbf k$. At the same time, Alice, as a sender, remains oblivious to Bob's input $y$ and the associated output $f(\mathbf{k}, y)$.
Finally, Alice sends $\mathcal A$ to Bob, and Bob computes $\mathcal A\cap \mathcal B$ to figure out the intersection of private sets $\mathcal X$ and $\mathcal Y$.

In this work, we integrate our QOT protocol into the optimized OPRF/PSI presented in~\cite{CCS:KKRT16}. The details are given in Section \ref{sec:OPRF}.
We conducted the PSI experiment on both simulated and real-world data. For real-world data, one bank provides blacklist accounts potentially involving telecommunication fraud among multiple institutions, while the other bank, acting as an inquirer, supplies active account data for the matching task. We want to find the intersection of two lists without revealing any further information.

\begin{figure}
    \centering
    \includegraphics[width=\linewidth]{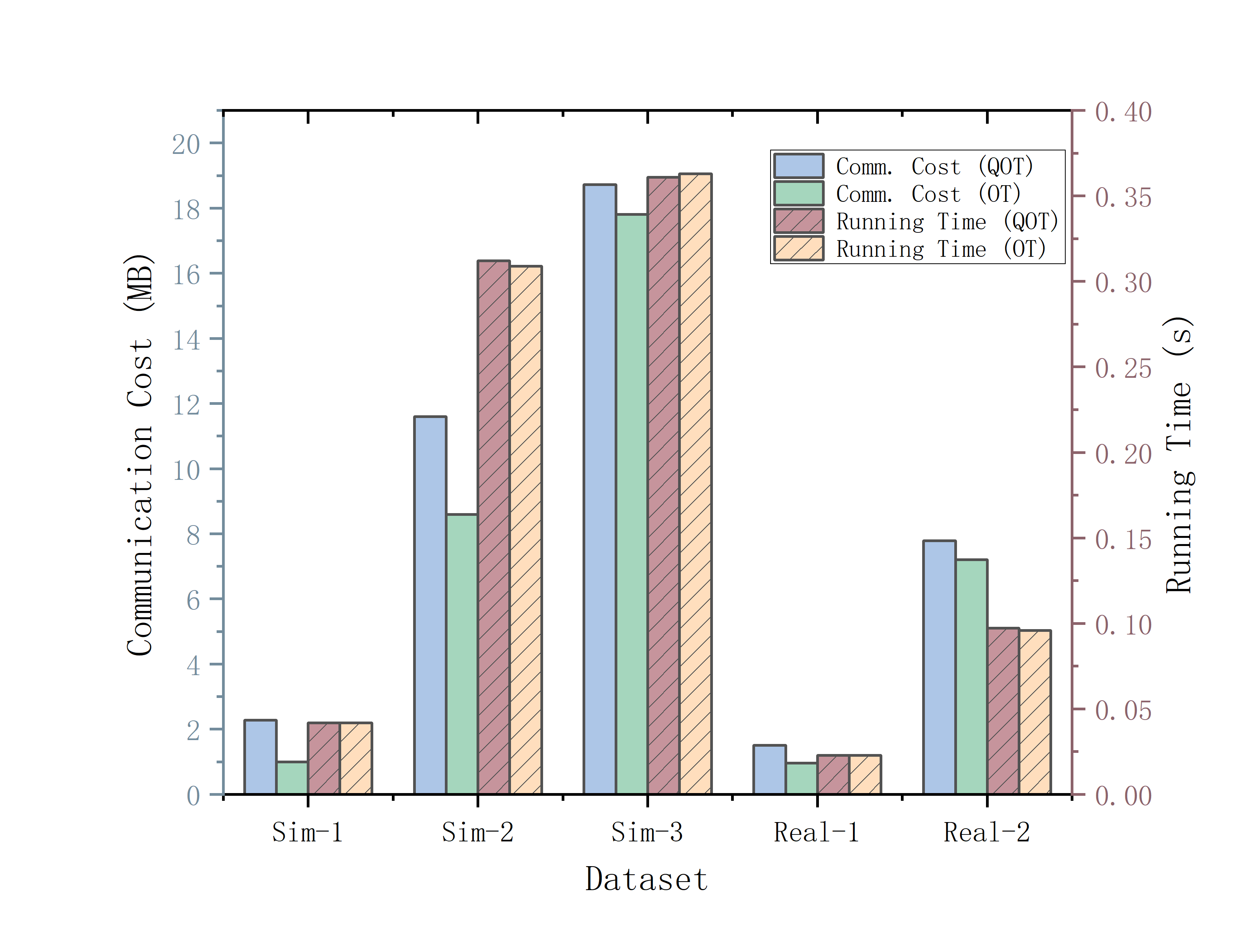}
    \caption{The comparison of communication cost and runtime between QOT-based and OT-based PSI experiments. Sim-1, Sim-2 and Sim-3 correspond to simulation data with different data sizes. Real-1 and Real-2 both correspond to real world data. From left to right, the number of queries in each PSI experiment are $10^3$, $10^3$, $10^5$, $10^4$ and $10^5$ respectively, and the number of elements in the private set in each experiment are $10^4$, $10^5$, $10^5$, $10^4$ and $10^4$ respectively. }
    \label{fig:performance}
\end{figure}

We apply our QOT protocol to solve this PSI problem, successfully identifying highly suspicious telecommunications fraud bank accounts.  To evaluate the performance of our protocol, we measure the communication cost and runtime during the PSI experiment, with the results illustrated in Figure~\ref{fig:performance}. To benchmark the performance of our QOT protocol, we also compare the communication cost and runtime of the same PSI experiment based on classical OT in Figure~\ref{fig:performance}.  Our results indicate that QOT incurs only slightly higher communication overhead compared to classical OT, while their runtime remains nearly the same. In exchange, our QOT protocol provides quantum security that classical OT cannot achieve.
Our experiment demonstrates the feasibility of the above QOT protocol.

\section{Discussion}

In this paper, we present an experimental implementation of a  protocol that is secure against any quantum attack. Unlike QOT protocols in the noisy storage model, our protocol does not rely on assumptions about noisy or limited quantum memory on the part of the adversary. Instead of restricting the capability of the adversary, our protocol employs a bit commitment scheme to guarantee security against delayed measurement attacks. The security of our QOT protocol is based on the well-founded security of bit commitment. Our protocol is secure against both classical and quantum attacks, within polynomial time.
The experimental demonstration uses an optical system adapted from a decoy state quantum key distribution (QKD) system, ensuring security against photon splitting attacks in practical single-photon sources.

While it is well known that secure multiparty computation (MPC) can be realized through OT~\cite{STOC:Kilian88}, no previous experiments have utilized quantum oblivious transfer (QOT) to solve real-world problems. In this paper, we are the first to apply our QOT protocol to address a practically significant private set intersection (PSI) problem in finance. Specifically, we demonstrate how two banks, one with a blacklist and the other with a list of active accounts, can compute the intersection of their private data sets without revealing any further information to each other. Our framework, when applied to other problem instances, enables secure MPC applications beyond PSI, including privacy-preserving machine learning~\cite{SP:MohZha17}  and anonymous voting systems~\cite{haenni2017cast}.

\section{Methods}

\subsection{Cryptographic assumption hierarchy}

Cryptographic primitives can be broadly categorized into three types: information-theoretic, symmetric-key, and public-key systems. The one-time pad is a prime example of information-theoretic security, providing unconditional security that can be rigorously proven mathematically. Symmetric-key systems, including symmetric encryption and hash functions, are based on complex, unstructured problems. In contrast, public-key systems depend on structured hard problems with algebraic representations, such as RSA. While the security of both symmetric-key and public-key systems is based on the presumed hardness of their underlying problems, symmetric-key systems are considered more foundational due to their reliance on weaker assumptions.

A further distinction exists between traditional cryptographic assumptions, like RSA and discrete logarithms, which are vulnerable to quantum attacks~\cite{FOCS:Shor94}, and other assumptions, such as lattice-based and code-based cryptography, which are not currently known to be susceptible to quantum algorithms~\cite{bernstein2017post}. Quantum information theory relies on established principles of quantum physics, which, if disproven, would indicate a discovery of new physics. Quantum key distribution is typically classified as information-theoretic, representing the highest level of security.
Oblivious transfer usually employs public-key cryptography for implementation. However, as noted in the literature~\cite{BBCS92,EC:GLSV21}, oblivious transfer protocols can also be constructed using only quantum information and symmetric-key encryption. This approach offers more reliable assurance without relying on strong computational assumptions.

\begin{figure*}[!ht]
    \begin{center}    
    \includegraphics[width=\textwidth]{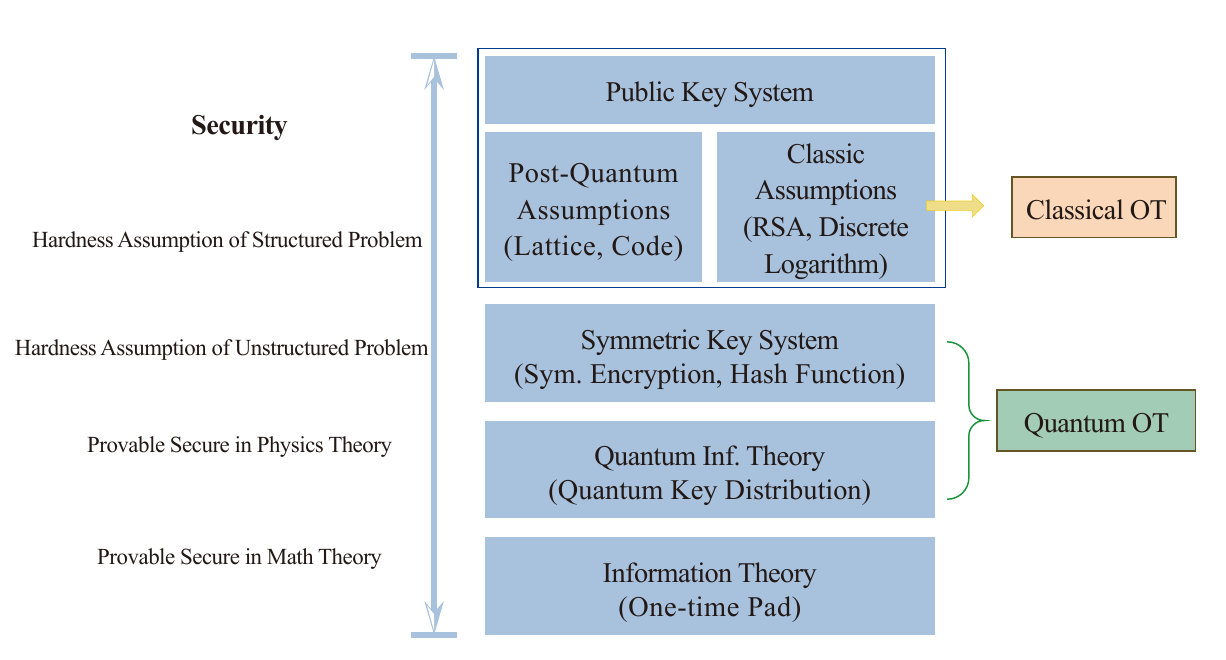}
    \caption{The hierarchy of cryptographic assumptions. From up to down, the assumptions become weaker, and the corresponding cryptographic protocols are regarded as more reliable.}
    \label{fig:hierarchy}
    \end{center}
\end{figure*}

As illustrated in Figure~\ref{fig:hierarchy}, our QOT protocol is based on weaker cryptographic assumptions compared to classical OT (or even post-quantum OT). Specifically, our protocol requires only quantum-secure bit commitment schemes, which depend on the existence of (quantumly hard) one-way permutations or collision-resistant hash functions—both categorized as unstructured symmetric mathematical problems. In contrast, post-quantum OT and public-key encryption typically rely on the conjectured hardness of lattice-based or code-based problems. This suggests that, while it is conceivable (though less likely) that the hardness of lattice and code problems could be disproven in the future, the existence of one-way permutations and collision-resistant hash functions is more likely to hold true. Moreover, the existence of (quantumly hard) one-way functions represents the minimal assumption necessary for meaningful cryptography beyond secure communication in the quantum realm.

\subsection{QOT protocol}
\label{sec:protocol}
Now we explain our QOT protocol step by step. Alice, as the sender, has three optical sources, one of which is for preparing signal states while the other two preparing decoy states. At each round $i$, Alice chooses two bits \(x_i \in \bool\) and \(\theta_i \in \bool\) independently at random. By randomly using one of her optical sources, Alice prepares a polarization-encoded single-photon state \(\ket{x_i}_{\theta_i}\) and then sends her encoded state to Bob. At the  receiver's side, Bob randomly chooses 
a bit \(\tilde\theta_i \in \bool\) and performs a projective measurement on the \(i\)-th received qubit in the basis \(\{\ket{0}_{\tilde{\theta}_i}, \ket{1}_{\tilde{\theta}_i}\}\), where \(\ket{0}_0 := \ket{0}\), \(\ket{1}_0 := \ket{1}\), \(\ket{0}_1 := \ket{+}\), and \(\ket{1}_1 := \ket{-}\). 

Due to imperfections in the photon source and dark counts, Bob may fail to detect any photons in certain rounds. Bob reports to Alice that rounds had no photon clicks at his detector. Alice then checks whether the ratio of Bob's reported rounds of photon loss falls within a reasonable range for each decoy state source, based on the estimated parameters of their quantum devices. If Bob's reported ratio is outside this reasonable range, Alice concludes that Bob must be cheating and abort the protocol. Otherwise, Alice proceeds as follows.

 The set of rounds where Alice sent signals while Bob reported photon clicks is denoted by $\mathcal M$ and the cardinality of $\mathcal M$ is $|\mathcal M|=n$.
To ensure that Bob measures all these $n$ qubits instead of storing some qubits in his quantum memory for measurements after Alice reveals her bases $\bm \theta:=\{\theta_i| i\in \mathcal M\}$, Bob must commit to all his values of \(\tilde{\bm{\theta}}:=\{\tilde{\theta}_i| i\in\mathcal M\}\) and \(\tilde{\bm{x}}:=\{\tilde{x}_i| i\in\mathcal M\}\). This step is the key difference between our QOT protocol from the one in the noisy storage model~\cite{erven2014}. This commitment can be achieved by employing a one-way collision-resistant hash function encoding the values of \(\tilde{\bm{x}}\) and \(\tilde{\bm{\theta}}\). For each \(i\in\mathcal M\), Bob chooses a random bit string \(\bm{r}_i\), uses a hash function \(h\), and sends the value \(c_i := h\left(\tilde{x}_i, \tilde{\theta}_i, \bm{r}_i\right)\) to Alice.
To verify Bob's commitment to his measurement bases, Alice randomly chooses a subset \(T \subset \mathcal M\) of size $|T|=\alpha n$, where $0<\alpha<1$, as a test set and reveals the committed values \(c_i\) for \(i \in T\). To open a commitment \(c_i\), Alice asks Bob to send her the values of \(\tilde{x}_i\), \(\tilde{\theta}_i\) and $\bm r_i$. Due to the collision resistance property of $h$, Alice can easily find out whether Bob changed his committed value. Then, Alice checks whether \(\tilde{x}_j = x_j\) holds true when \(\tilde{\theta}_j = \theta_j\) for \(j \in T\) with a ratio at least $\beta\in (1/2,1)$. If it holds, then Alice passes this verification and continues; otherwise, she aborts. After that, both Alice and Bob discard the test qubits in the set \(T\), and Alice sends all her bases \(\{\theta_i| i\in\mathcal M\setminus T\}\) to Bob.

 When $\theta_i=\tilde{\theta}_i$, the bits $x_i$ and $\tilde{x}_i$ match with a high rate, but appear independently random otherwise. While Bob knows which ones correspond to $\theta_i=\tilde{\theta}_i$, Alice remains unaware of this information.
Bob randomly chooses a choice bit \(c \in \{0, 1\}\) and divides the set \( \mathcal M \setminus T\) into two subsets: a good set \(I_c := \{i \in \mathcal M\setminus T \mid \theta_i = \tilde{\theta}_i\}\) and a bad set \(I_{1-c} := \{i \in \mathcal M\setminus T\mid \theta_i \neq \tilde{\theta}_i\}\). 
The bit strings $\bm x$ is then divided into two $\bm x_{I_c}$ and $\bm x_{I_{1-c}}$ on $I_c$ and $I_{1-c}$ respectively. Similarly, $\tilde{\bm x}$ is divided into  $\tilde{\bm x}_{I_c}$ and $\tilde{\bm x}_{I_{1-c}}$.
For simplicity, we suppose $|I_c|=|I_{1-c}|$. In the case where $I_c$ and $I_{1-c}$ have different lengths, we truncate the longer one to the length of the other.
Bob sends both sets \(I_0\) and \(I_1\) to Alice, without informing her which one corresponds to the good one. 

Due to transmission errors as well as measurement errors, even for \(i\in I_c\), there is still some chance that \(x_i \neq \tilde{x}_i\). To reduce such bit-flip errors in practice, Alice and Bob employ a classical error correction code to construct noiseless shared random secret keys. For example, they can use Bose-Chaudhuri-Hocquenghem (BCH) codes~\cite{bose1960class} that are characterized by their ability to correct multiple random errors. 
Alice generates random secret keys \(\tilde{\bm r}_0\) and \(\tilde{\bm r}_1\) of the same length and uses an error correction code to encode them. Then she sends to Bob \(\bm{y}_0 := \text{Encode}(\tilde{\bm r}_0) \oplus \bm{x}_{I_0}\) and \(\bm{y}_1 := \text{Encode}(\tilde{\bm r}_1) \oplus \bm{x}_{I_1}\), where Encode denotes the encoding function of the error correction code and $\oplus$ denotes addition modulo two.
After receiving $\bm y_0$ and $\bm y_1$, Bob uses his bit strings \(\tilde{\bm{x}}_{I_c}\) to calculate \(\text{Decode}\left(\bm{y}_c\oplus \tilde{\bm{x}}_{I_c}\right)\), where Decode denotes the decoding function of the error correction code.
When the number of bit flip errors $\sum_{i\in I_c} |x_i\oplus \tilde{x}_i|$ is below a certain error threshold, Bob can perfectly recover $\tilde{\bm r}_c=\text{Decode}(\text{Encode}(\tilde{\bm r}_c) \oplus \bm{x}_{I_c} \oplus \tilde{\bm x}_{I_c})$. On the other hand, as $\bm{x}_{1-c}$ and $\tilde{\bm x}_{1-c}$ are independently random, Bob can rarely learn information about $\tilde{\bm r}_{1-c}$.

Up to now, Alice has obliviously transferred random keys $\tilde{\bm r}_0$ and $\tilde{\bm r}_1$ to Bob, with Bob only receiving $\tilde{\bm r}_c$.
Alice can now use these obliviously transferred random secret keys to accomplish the OT protocol. 
However, the probability that Bob can learn information about $\tilde{\bm r}_{1-c}$ is nonzero. To minimize the information Bob can retrieve about $\tilde{\bm r}_{1-c}$, we use privacy amplification, ensuring that Bob ultimately learns nothing about the secret key. By using two universal hash functions $f_0$ and $f_1$ agreed upon by both Alice and Bob, they obtain bit strings $f_0(\tilde{\bm r}_0)$ and $f_1(\tilde{\bm r}_1)$ of shorter length $\lambda$ from the original random keys $\tilde{\bm r}_0$ and $\tilde{\bm r}_1$. 
Suppose Alice has two pieces of messages \(\bm{m}_0\) and \(\bm{m}_1\) of length $\lambda$. 
Using the secret keys \(f_0(\tilde{\bm r}_0)\) and \(f_1(\tilde{\bm r}_1)\) as one-time pads, Alice sends Bob her encrypted messages \(\bm{z}_0 := \bm{m}_0 \oplus f_0(\tilde{\bm r}_0)\) and \(\bm{z}_1 := \bm{m}_1 \oplus f_1(\tilde{\bm r}_1)\). 
Hence, Bob uses $f_c(\tilde{\bm r}_c)$ to recover the message \(\bm{m}_c = \bm{z}_c \oplus f_c(\tilde{\bm r}_c)\) correctly.
At the same time, since Alice does not know the value of \(c\), she is oblivious to which of \(\bm{m}_0\) and \(\bm{m}_1\) has been successfully sent to Bob throughout the entire protocol.

\subsection{Correctness and security analysis}
\label{sec:CandSanalysis}

\begin{table}[]
    \begin{tabular}{@{}cccc@{}}
        \toprule
        physical meaning of parameters & values \\
        \midrule
        length of message  &  $\lambda =256$\\ 
        number of single photon states received by Bob  &  $n=2044$\\ 
       ratio of test bits & $\alpha=1/2$ \\ 
       minimum acceptable ratio of bit matching &  $\beta=0.95$ \\ 
       error correction code length  & $N=511$ \\ 
        maximum number of correctable bit flip errors & $t=30$ \\
       \bottomrule
    \end{tabular}
    \centering
    \vspace{10pt}
    \caption{The parameter choices used in our experimental implementation}
    \label{tab:parameters}
\end{table}

Assuming independent and identical noise in each round, let the overall bit flip error rate be $p_e$. 
The correctness of our protocol relies on two parts. First, Bob must pass the check in Steps 3-5. Second, the number of matching bits in set $I_C$, i.e.\ $|\{i\in I_C|\tilde x_i=x_i\}|$, must be at least $N-t$ to ensure that the message can be decoded in Step 9. Therefore, the failure probability is 
$ p_{\textrm{failed}}\leq  p_{\textrm{fpass}} + p_{\textrm{fcorrect}}$, 
where 
\begin{align*}
p_{\textrm{fpass}} =   & \sum_{i=0}^{\beta\alpha n}\binom{\alpha n}{i}p_e^{\alpha n -i}(1-p_e)^i, \textrm{ and}  \\
p_{\textrm{fcorrect}} =  & 1  -\frac{1}{2^{n-\alpha n}}\sum_{i=N+1}^{n-\alpha n} \left(\binom{n-\alpha n}{i}\cdot \left(\sum_{j=N-t}^N\binom{N}{j}p_e^{N-j}(1-p_e)^j\right)\right)\\
&-\frac{1}{2^{n-\alpha n}}\sum_{i=0}^N \left(\binom{n-\alpha n}{i}\cdot \left(\sum_{j=0}^i\binom{i}{j}p_e^{i-j}(1-p_e)^j\cdot \frac{1}{2^{N-i}}\sum_{k=N-t-j}^{N-i}\binom{N-i}{k}\right)\right).
\end{align*}
Using the parameters listed in Table~\ref{tab:parameters}, we find that the probability is at most $2.3 \times 10^{-9}$,
which is quite close to zero.

Next, let us consider the security of our protocol against a malicious Bob. Bob can attempt to steal information from Alice by delaying his measurement in step 2, until Alice reveals the correct basis in step 6. However, he must first bypass the checks in steps 3-5.
Assuming \( p_e = 0 \), which is favorable for Bob, he can delay the measurement of \( s_1 \) qubits (where \( s_1 \leq \alpha n \)) in step 2 until Alice reveals her bases and guess the measurement outcomes of $s_2$ qubits in the set $I_{1-c}$. The probability that he bypasses the first check is
\begin{align*}
p_{\textrm{bypass},s_1} & = \frac{1}{\binom{n}{\alpha n}} \cdot \sum_{i=0}^{s_1} \binom{s_1}{i} \binom{n-s_1}{\alpha n -i}\Pr[\textrm{check passed}],
\end{align*}
where
\begin{align*}
    \Pr[\textrm{check passed}] = \begin{cases}
    1 &\text{if } \alpha n -i \geq \beta\alpha n, \\
    \sum_{j=\lceil i-(1-\beta)\alpha n  \rceil}^i \binom{i}{j} \frac{1}{2^j}\cdot \frac{1}{2^{i-j}} & \text{otherwise}.
\end{cases}
\end{align*}
Once Bob bypasses the first check, he must obtain the correct values for more than \(2(N-t) \) qubits to acquire both messages. He can get $s_1$ bits from the associated qubits delayed for measurements and another $s_2$ bits from guessing. The probability of this happening is
\begin{equation*}
    p_{\text{cheat},s_1,s_2} = \frac{1}{2^{(1-\alpha)n}}\sum_{i=2(N-t)-s_1-s_2}^{(1-\alpha) n}\binom{(1-\alpha)n}{i}. 
\end{equation*}
By defining the cheating cost as the expected time for an attacker to successfully deceive once, Bob's final cheating cost is 
$$ {\text{cost}}_{\text{cheat}} = \min_{0\leq s_1\leq \alpha, 0\leq s_2\leq 2(N-t)-s_1} 
\frac{2^{s_2}}{p_{\textrm{bypass},s_1}\cdot p_{\text{cheat},s_1,s_2}} \geq 2.8 \times 10^{12}.$$

\subsection{Potential Optimization}

Our QOT protocol, as well as OT protocol, has some potential optimization methods to reduce the computation and communication cost.

\paragraph{OT Extension.} 
Oblivious transfer (OT) extension enables us to extend the number of OTs performed using a small set of base OTs. Ishai et al.~\cite{C:IKNP03} introduced a framework for OT extension within the Random Oracle Model (ROM), which only
 makes use of cheap symmetric key primitives. Following studies have built upon the Ishai et al. ~\cite{C:IKNP03} framework, subsequent
 works enhancing it to increase security level with better efficiency~\cite{C:KolKum13,C:KelOrsSch15}.

\paragraph{Preprocessing.} Beaver triples~\cite{C:Beaver91b} are pre-shared correlated randomness and are widely used in secure multiparty computation. Beaver triple can be pre-generated by OT in the offline phase. In the online phase, parties can execute MPC protocol by beaver triples without executing oblivious transfer.


\subsection{Private set intersection from oblivious pseudorandom function}
\label{sec:OPRF}

\begin{figure*}[!ht]
\begin{nfprot}{OPRF}
\noindent \textbf{Input of $\mathcal{R}$:} $m$ selection strings $r = (r_1, \ldots, r_m)$, $r_i \in \{0,1\}^*$. \\
\textbf{Parameters:}
\begin{itemize}
    \item A $(\kappa, \epsilon)$-PRC family $\mathcal{C}$ with output length $k = k(\kappa)$.
    \item A $\kappa$-Hamming correlation-robust $H: [m] \times \{0,1\}^k \rightarrow \{0,1\}^v$.
    \item An ideal $\text{OT}^k_m$ primitive.
\end{itemize}

\noindent \textbf{Protocol:}
\begin{enumerate}
    \item $\mathcal{S}$ chooses a random $C \leftarrow \mathcal{C}$ and sends it to $\mathcal{R}$.
    \item $\mathcal{S}$ chooses $s \leftarrow \{0,1\}^k$ at random. Let $s_i$ denote the $i$-th bit of $s$.
    \item $\mathcal{R}$ forms $m \times k$ matrices $T_0, T_1$ in the following way:
    \begin{itemize}
        \item For $j \in [m]$, choose $t_{0,j} \leftarrow \{0,1\}^k$ and set $t_{1,j} = C(r_j) \oplus t_{0,j}$.
    \end{itemize}
    Let $t_i^0, t_i^1$ denote the $i$-th column of matrices $T_0, T_1$ respectively.
    \item $\mathcal{S}$ and $\mathcal{R}$ interact with $\text{OT}^k_m$ in the following way:
    \begin{itemize}
        \item $\mathcal{S}$ acts as receiver with input $\{s_i\}_{i \in [k]}$.
        \item $\mathcal{R}$ acts as sender with input $\{t_i^0, t_i^1\}_{i \in [k]}$.
        \item $\mathcal{S}$ receives output $\{q^i\}_{i \in [k]}$.
    \end{itemize}
    $\mathcal{S}$ forms $m \times k$ matrix $Q$ such that the $i$-th column of $Q$ is the vector $q^i$. (Note $q^i = t_{s_i}^i$.) Let $q_j$ denote the $j$-th row of $Q$. Note, $q_j = ((t_{0,j} \oplus t_{1,j}) \cdot s) \oplus t_{0,j}$. Simplifying, $q_j = t_{0,j} \oplus (C(r_j) \cdot s)$.
    \item For $j \in [m]$, $\mathcal{S}$ outputs the PRF seed $((C, s), (j, q_j))$.
    \item For $j \in [m]$, $\mathcal{R}$ outputs relaxed PRF output $(C, j, t_{0,j})$.
\end{enumerate}
\end{nfprot}
\vspace{10pt}
\caption{Oblivious PRF Protocol}
\label{fig:oprf}
\end{figure*}

\begin{figure*}[!ht]
\begin{nfprot}{PSI}
\noindent \textbf{Parameters:} Alice has input $X$; Bob has input $Y$, with $|X| = |Y| = n$. $s$ is an upper bound on the stash size for Cuckoo hashing.
\begin{enumerate}
    \item Bob specifies random hash functions $h_1, h_2, h_3 : \{0, 1\}^* \to [1.2n]$ and tells them to Alice.
    
    \item Bob assigns his items $Y$ into $1.2n$ bins using Cuckoo hashing. Let Bob keep track of $z(y)$ for each $y$ so that: if $z(y) = \perp$ then $y$ is in the stash; otherwise $y$ is in bin $h_{z(y)}(y)$. Arrange the items in the stash in an arbitrary order.
    
    Bob selects OPRF inputs as follows: for $i \in [1.2n]$, if bin $i$ is empty, then set $r_i$ to a dummy value; otherwise if $y$ is in bin $i$ then set $r_i = y \| z(y)$. For $i \in [s]$, if position $i$ in the stash is $y$, then set $r_i = y$; otherwise set $r_i$ to a dummy value.
    
    \item The parties invoke $1.2n + s$ OPRF instances, with Bob the receiver with inputs $(r_1, \dots, r_{1.2n+s})$. Alice receives $(k_1, \dots, k_{1.2n+s})$ and Bob receives $F(k_i, r_i)$ for all $i$.
    
    \item Alice computes:
    \[
    H_i = \{ F(k_{h_i(x)}, x \| i) \mid x \in X \}, \text{ for } i \in \{1, 2, 3\}
    \]
    \[
    S_j = \{ F(k_{1.2n+j}, x) \mid x \in X \}, \text{ for } j \in \{1, \dots, s\}
    \]
    and sends a permutation of each set to Bob.
    
    \item Bob initializes an empty set $\mathcal{O}$ and does the following for $y \in Y$: If $z(y) = \perp$ and $y$ is at position $j$ in the stash and $F(k_{1.2n+j}, y) \in S_j$, then Bob adds $y$ to $\mathcal{O}$. If $z(y) \neq \perp$ and $F(k_{h_{z(y)}(y)}, y \| z(y)) \in H_{z(y)}$ then Bob adds $y$ to $\mathcal{O}$.
    
    \item Bob sends $\mathcal{O}$ to Alice and both parties output $\mathcal{O}$.
\end{enumerate}
\end{nfprot}
\vspace{10pt}
\caption{Private Set Intersection Protocol}
\label{fig:psi}
\end{figure*}

The BaRK-OPRF protocol~\cite{CCS:KKRT16} is an efficient implementation of OPRF.  This is accomplished through a batchable approach, where multiple evaluations can be handled efficiently in a single protocol execution.

The protocol works as follows. The sender generates a secret key for the PRF and shares the key in a secure, oblivious manner using a cryptographic protocol like oblivious transfer.
The receiver prepares their input values for which they want to evaluate the PRF.
Through an interactive process, the sender and receiver engage in a protocol where the receiver obtains the PRF evaluations on their inputs without the sender learning the inputs or the outputs.
 The protocol is designed to handle multiple inputs efficiently in a batch, reducing computational overhead and improving performance.
The protocol ensures that the receiver only learns the PRF values for their specific inputs, and the sender remains unaware of which inputs were chosen and the corresponding outputs. The detailed algorithm is summarized in the box of Protocol OPRF.

Based on the BaRK-OPRF protocol, ~\cite{CCS:KKRT16} presents an improved PSI protocol. The detailed algorithm is listed in Fig.~\ref{fig:oprf} and Fig.~\ref{fig:psi}.

\backmatter

\section{Declaration}

\subsection*{DATA AVAILABILITY}
The data that support the findings of this study are available from the corresponding author upon reasonable request.

\subsection*{MATERIALS AVAILABILITY}
The materials that support the findings of this study are available from the corresponding author upon reasonable request.

\subsection*{SOFTWARE AVAILABILITY}
The code that support the findings of this study are available from the corresponding author upon reasonable request.

\subsection*{ACKNOWLEDGMENTS} 
The authors acknowledge Chengfang Jinke and Minfeng Bank for providing us with the data and the application scenario that facilitate the research. 

\subsection*{AUTHOR CONTRIBUTIONS}
Kai-Yi Zhang, Ya-Dong Wu and Yu Yu designed research; 
 An-Jing Huang, Kun Tu, Ming-Han Li, Chi Zhang and Wei Qi performed the experiment. All authors discussed the results and reviewed the manuscript.

\subsection*{COMPETING INTERESTS}
The authors declare no competing interests.



\input{main.bbl}

\end{document}

%% file: main.bbl

%% file: main.bbl
\begin{thebibliography}{38}
\ifx \bisbn   \undefined \def \bisbn  #1{ISBN #1}\fi
\ifx \binits  \undefined \def \binits#1{#1}\fi
\ifx \bauthor  \undefined \def \bauthor#1{#1}\fi
\ifx \batitle  \undefined \def \batitle#1{#1}\fi
\ifx \bjtitle  \undefined \def \bjtitle#1{#1}\fi
\ifx \bvolume  \undefined \def \bvolume#1{\textbf{#1}}\fi
\ifx \byear  \undefined \def \byear#1{#1}\fi
\ifx \bissue  \undefined \def \bissue#1{#1}\fi
\ifx \bfpage  \undefined \def \bfpage#1{#1}\fi
\ifx \blpage  \undefined \def \blpage #1{#1}\fi
\ifx \burl  \undefined \def \burl#1{\textsf{#1}}\fi
\ifx \doiurl  \undefined \def \doiurl#1{\url{https://doi.org/#1}}\fi
\ifx \betal  \undefined \def \betal{\textit{et al.}}\fi
\ifx \binstitute  \undefined \def \binstitute#1{#1}\fi
\ifx \binstitutionaled  \undefined \def \binstitutionaled#1{#1}\fi
\ifx \bctitle  \undefined \def \bctitle#1{#1}\fi
\ifx \beditor  \undefined \def \beditor#1{#1}\fi
\ifx \bpublisher  \undefined \def \bpublisher#1{#1}\fi
\ifx \bbtitle  \undefined \def \bbtitle#1{#1}\fi
\ifx \bedition  \undefined \def \bedition#1{#1}\fi
\ifx \bseriesno  \undefined \def \bseriesno#1{#1}\fi
\ifx \blocation  \undefined \def \blocation#1{#1}\fi
\ifx \bsertitle  \undefined \def \bsertitle#1{#1}\fi
\ifx \bsnm \undefined \def \bsnm#1{#1}\fi
\ifx \bsuffix \undefined \def \bsuffix#1{#1}\fi
\ifx \bparticle \undefined \def \bparticle#1{#1}\fi
\ifx \barticle \undefined \def \barticle#1{#1}\fi
\bibcommenthead
\ifx \bconfdate \undefined \def \bconfdate #1{#1}\fi
\ifx \botherref \undefined \def \botherref #1{#1}\fi
\ifx \url \undefined \def \url#1{\textsf{#1}}\fi
\ifx \bchapter \undefined \def \bchapter#1{#1}\fi
\ifx \bbook \undefined \def \bbook#1{#1}\fi
\ifx \bcomment \undefined \def \bcomment#1{#1}\fi
\ifx \oauthor \undefined \def \oauthor#1{#1}\fi
\ifx \citeauthoryear \undefined \def \citeauthoryear#1{#1}\fi
\ifx \endbibitem  \undefined \def \endbibitem {}\fi
\ifx \bconflocation  \undefined \def \bconflocation#1{#1}\fi
\ifx \arxivurl  \undefined \def \arxivurl#1{\textsf{#1}}\fi
\csname PreBibitemsHook\endcsname

\bibitem[\protect\citeauthoryear{Yao}{1982}]{FOCS:Yao82b}
\begin{bchapter}
\bauthor{\bsnm{Yao}, \binits{A.C.-C.}}:
\bctitle{Protocols for secure computations (extended abstract)}.
In: \bbtitle{23rd Annual Symposium on Foundations of Computer Science},
pp. \bfpage{160}--\blpage{164}.
\bpublisher{{IEEE} Computer Society Press},
\blocation{Chicago, Illinois}
(\byear{1982}).
\doiurl{10.1109/SFCS.1982.38}
\end{bchapter}
\endbibitem

\bibitem[\protect\citeauthoryear{Yao}{1986}]{FOCS:Yao86}
\begin{bchapter}
\bauthor{\bsnm{Yao}, \binits{A.C.-C.}}:
\bctitle{How to generate and exchange secrets (extended abstract)}.
In: \bbtitle{27th Annual Symposium on Foundations of Computer Science},
pp. \bfpage{162}--\blpage{167}.
\bpublisher{{IEEE} Computer Society Press},
\blocation{Toronto, Ontario, Canada}
(\byear{1986}).
\doiurl{10.1109/SFCS.1986.25}
\end{bchapter}
\endbibitem

\bibitem[\protect\citeauthoryear{Goldreich et~al.}{1987}]{STOC:GolMicWig87}
\begin{bchapter}
\bauthor{\bsnm{Goldreich}, \binits{O.}},
\bauthor{\bsnm{Micali}, \binits{S.}},
\bauthor{\bsnm{Wigderson}, \binits{A.}}:
\bctitle{How to play any mental game or {A} completeness theorem for protocols with honest majority}.
In: \beditor{\bsnm{Aho}, \binits{A.}} (ed.)
\bbtitle{19th Annual {ACM} Symposium on Theory of Computing},
pp. \bfpage{218}--\blpage{229}.
\bpublisher{{ACM} Press},
\blocation{New York City, NY, USA}
(\byear{1987}).
\doiurl{10.1145/28395.28420}
\end{bchapter}
\endbibitem

\bibitem[\protect\citeauthoryear{Mohassel and Zhang}{2017}]{SP:MohZha17}
\begin{bchapter}
\bauthor{\bsnm{Mohassel}, \binits{P.}},
\bauthor{\bsnm{Zhang}, \binits{Y.}}:
\bctitle{{SecureML}: {A} system for scalable privacy-preserving machine learning}.
In: \bbtitle{2017 {IEEE} Symposium on Security and Privacy},
pp. \bfpage{19}--\blpage{38}.
\bpublisher{{IEEE} Computer Society Press},
\blocation{San Jose, CA, USA}
(\byear{2017}).
\doiurl{10.1109/SP.2017.12}
\end{bchapter}
\endbibitem

\bibitem[\protect\citeauthoryear{Cho et~al.}{2018}]{cho2018secure}
\begin{barticle}
\bauthor{\bsnm{Cho}, \binits{H.}},
\bauthor{\bsnm{Wu}, \binits{D.J.}},
\bauthor{\bsnm{Berger}, \binits{B.}}:
\batitle{Secure genome-wide association analysis using multiparty computation}.
\bjtitle{Nature biotechnology}
\bvolume{36}(\bissue{6}),
\bfpage{547}--\blpage{551}
(\byear{2018})
\end{barticle}
\endbibitem

\bibitem[\protect\citeauthoryear{Kilian}{1988}]{STOC:Kilian88}
\begin{bchapter}
\bauthor{\bsnm{Kilian}, \binits{J.}}:
\bctitle{Founding cryptography on oblivious transfer}.
In: \bbtitle{20th Annual {ACM} Symposium on Theory of Computing},
pp. \bfpage{20}--\blpage{31}.
\bpublisher{{ACM} Press},
\blocation{Chicago, IL, USA}
(\byear{1988}).
\doiurl{10.1145/62212.62215}
\end{bchapter}
\endbibitem

\bibitem[\protect\citeauthoryear{Even et~al.}{1982}]{C:EveGolLem82}
\begin{bchapter}
\bauthor{\bsnm{Even}, \binits{S.}},
\bauthor{\bsnm{Goldreich}, \binits{O.}},
\bauthor{\bsnm{Lempel}, \binits{A.}}:
\bctitle{A randomized protocol for signing contracts}.
In: \beditor{\bsnm{Chaum}, \binits{D.}},
\beditor{\bsnm{Rivest}, \binits{R.L.}},
\beditor{\bsnm{Sherman}, \binits{A.T.}} (eds.)
\bbtitle{Advances in Cryptology -- {CRYPTO}'82},
pp. \bfpage{205}--\blpage{210}.
\bpublisher{Plenum Press, New York, USA},
\blocation{Santa Barbara, CA, USA}
(\byear{1982})
\end{bchapter}
\endbibitem

\bibitem[\protect\citeauthoryear{Naor and Pinkas}{2001}]{SODA:NaoPin01}
\begin{bchapter}
\bauthor{\bsnm{Naor}, \binits{M.}},
\bauthor{\bsnm{Pinkas}, \binits{B.}}:
\bctitle{Efficient oblivious transfer protocols}.
In: \beditor{\bsnm{Kosaraju}, \binits{S.R.}} (ed.)
\bbtitle{12th Annual {ACM}-{SIAM} Symposium on Discrete Algorithms},
pp. \bfpage{448}--\blpage{457}.
\bpublisher{{ACM-SIAM}},
\blocation{Washington, DC, USA}
(\byear{2001})
\end{bchapter}
\endbibitem

\bibitem[\protect\citeauthoryear{Santos et~al.}{2022}]{santos2022quantum}
\begin{barticle}
\bauthor{\bsnm{Santos}, \binits{M.B.}},
\bauthor{\bsnm{Mateus}, \binits{P.}},
\bauthor{\bsnm{Pinto}, \binits{A.N.}}:
\batitle{Quantum oblivious transfer: a short review}.
\bjtitle{Entropy}
\bvolume{24}(\bissue{7}),
\bfpage{945}
(\byear{2022})
\end{barticle}
\endbibitem

\bibitem[\protect\citeauthoryear{Mayers}{1997}]{May97}
\begin{botherref}
\oauthor{\bsnm{Mayers}, \binits{D.}}:
Unconditionally secure quantum bit commitment is impossible
\textbf{78}(17),
3414--3417
(1997)
\doiurl{10.1103/PhysRevLett.78.3414}
{\href{https://arxiv.org/abs/9605044}{{arXiv:9605044}}}
{[quant-ph]}
\end{botherref}
\endbibitem

\bibitem[\protect\citeauthoryear{Lo}{1997}]{Lo97}
\begin{botherref}
\oauthor{\bsnm{Lo}, \binits{H.-K.}}:
Insecurity of quantum secure computations
\textbf{56}(2),
1154--1162
(1997)
\doiurl{10.1103/PhysRevA.56.1154}
\end{botherref}
\endbibitem

\bibitem[\protect\citeauthoryear{Bennett et~al.}{2001}]{BBCS92}
\begin{bchapter}
\bauthor{\bsnm{Bennett}, \binits{C.H.}},
\bauthor{\bsnm{Brassard}, \binits{G.}},
\bauthor{\bsnm{Cr\'{e}peau}, \binits{C.}},
\bauthor{\bsnm{Skubiszewska}, \binits{M.-H.}}:
\bctitle{Practical quantum oblivious transfer},
pp. \bfpage{351}--\blpage{366}
(\byear{2001}).
\doiurl{10.1007/3-540-46766-1\_29}
\end{bchapter}
\endbibitem

\bibitem[\protect\citeauthoryear{Wehner et~al.}{2008}]{PhysRevLett.100.220502}
\begin{barticle}
\bauthor{\bsnm{Wehner}, \binits{S.}},
\bauthor{\bsnm{Schaffner}, \binits{C.}},
\bauthor{\bsnm{Terhal}, \binits{B.M.}}:
\batitle{Cryptography from noisy storage}.
\bjtitle{Phys. Rev. Lett.}
\bvolume{100},
\bfpage{220502}
(\byear{2008})
\doiurl{10.1103/PhysRevLett.100.220502}
\end{barticle}
\endbibitem

\bibitem[\protect\citeauthoryear{Wehner et~al.}{2010}]{PhysRevA.81.052336}
\begin{barticle}
\bauthor{\bsnm{Wehner}, \binits{S.}},
\bauthor{\bsnm{Curty}, \binits{M.}},
\bauthor{\bsnm{Schaffner}, \binits{C.}},
\bauthor{\bsnm{Lo}, \binits{H.-K.}}:
\batitle{Implementation of two-party protocols in the noisy-storage model}.
\bjtitle{Phys. Rev. A}
\bvolume{81},
\bfpage{052336}
(\byear{2010})
\doiurl{10.1103/PhysRevA.81.052336}
\end{barticle}
\endbibitem

\bibitem[\protect\citeauthoryear{Schaffner}{2010}]{PhysRevA.82.032308}
\begin{barticle}
\bauthor{\bsnm{Schaffner}, \binits{C.}}:
\batitle{Simple protocols for oblivious transfer and secure identification in the noisy-quantum-storage model}.
\bjtitle{Phys. Rev. A}
\bvolume{82},
\bfpage{032308}
(\byear{2010})
\doiurl{10.1103/PhysRevA.82.032308}
\end{barticle}
\endbibitem

\bibitem[\protect\citeauthoryear{Erven et~al.}{2014}]{erven2014}
\begin{barticle}
\bauthor{\bsnm{Erven}, \binits{C.}},
\bauthor{\bsnm{Ng}, \binits{N.}},
\bauthor{\bsnm{Gigov}, \binits{N.}},
\bauthor{\bsnm{Laflamme}, \binits{R.}},
\bauthor{\bsnm{Wehner}, \binits{S.}},
\bauthor{\bsnm{Weihs}, \binits{G.}}:
\batitle{An experimental implementation of oblivious transfer in the noisy storage model}.
\bjtitle{Nat. Commun.}
\bvolume{5}(\bissue{1}),
\bfpage{3418}
(\byear{2014})
\end{barticle}
\endbibitem

\bibitem[\protect\citeauthoryear{Zhu et~al.}{2022}]{zhu2022demand}
\begin{barticle}
\bauthor{\bsnm{Zhu}, \binits{T.-X.}},
\bauthor{\bsnm{Liu}, \binits{C.}},
\bauthor{\bsnm{Jin}, \binits{M.}},
\bauthor{\bsnm{Su}, \binits{M.-X.}},
\bauthor{\bsnm{Liu}, \binits{Y.-P.}},
\bauthor{\bsnm{Li}, \binits{W.-J.}},
\bauthor{\bsnm{Ye}, \binits{Y.}},
\bauthor{\bsnm{Zhou}, \binits{Z.-Q.}},
\bauthor{\bsnm{Li}, \binits{C.-F.}},
\bauthor{\bsnm{Guo}, \binits{G.-C.}}:
\batitle{On-demand integrated quantum memory for polarization qubits}.
\bjtitle{Physical Review Letters}
\bvolume{128}(\bissue{18}),
\bfpage{180501}
(\byear{2022})
\end{barticle}
\endbibitem

\bibitem[\protect\citeauthoryear{Naor}{1991}]{JC:Naor91}
\begin{barticle}
\bauthor{\bsnm{Naor}, \binits{M.}}:
\batitle{Bit commitment using pseudorandomness}.
\bjtitle{Journal of Cryptology}
\bvolume{4}(\bissue{2}),
\bfpage{151}--\blpage{158}
(\byear{1991})
\doiurl{10.1007/BF00196774}
\end{barticle}
\endbibitem

\bibitem[\protect\citeauthoryear{Furrer et~al.}{2018}]{furrer2018}
\begin{barticle}
\bauthor{\bsnm{Furrer}, \binits{F.}},
\bauthor{\bsnm{Gehring}, \binits{T.}},
\bauthor{\bsnm{Schaffner}, \binits{C.}},
\bauthor{\bsnm{Pacher}, \binits{C.}},
\bauthor{\bsnm{Schnabel}, \binits{R.}},
\bauthor{\bsnm{Wehner}, \binits{S.}}:
\batitle{Continuous-variable protocol for oblivious transfer in the noisy-storage model}.
\bjtitle{Nature communications}
\bvolume{9}(\bissue{1}),
\bfpage{1450}
(\byear{2018})
\end{barticle}
\endbibitem

\bibitem[\protect\citeauthoryear{Amiri et~al.}{2021}]{PRXQuantum2021}
\begin{barticle}
\bauthor{\bsnm{Amiri}, \binits{R.}},
\bauthor{\bsnm{St{\'a}rek}, \binits{R.}},
\bauthor{\bsnm{Reichmuth}, \binits{D.}},
\bauthor{\bsnm{Puthoor}, \binits{I.V.}},
\bauthor{\bsnm{Mi{\v{c}}uda}, \binits{M.}},
\bauthor{\bsnm{Mi{\v{s}}ta}, \binits{L.} \bsuffix{Jr}},
\bauthor{\bsnm{Du{\v{s}}ek}, \binits{M.}},
\bauthor{\bsnm{Wallden}, \binits{P.}},
\bauthor{\bsnm{Andersson}, \binits{E.}}:
\batitle{Imperfect 1-out-of-2 quantum oblivious transfer: Bounds, a protocol, and its experimental implementation}.
\bjtitle{PRX Quantum}
\bvolume{2},
\bfpage{010335}
(\byear{2021})
\doiurl{10.1103/PRXQuantum.2.010335}
\end{barticle}
\endbibitem

\bibitem[\protect\citeauthoryear{Stroh et~al.}{2023}]{PRXQuantum2023}
\begin{barticle}
\bauthor{\bsnm{Stroh}, \binits{L.}},
\bauthor{\bsnm{Horov{\'a}}, \binits{N.}},
\bauthor{\bsnm{St{\'a}rek}, \binits{R.}},
\bauthor{\bsnm{Puthoor}, \binits{I.V.}},
\bauthor{\bsnm{Mi{\v{c}}uda}, \binits{M.}},
\bauthor{\bsnm{Du{\v{s}}ek}, \binits{M.}},
\bauthor{\bsnm{Andersson}, \binits{E.}}:
\batitle{Noninteractive xor quantum oblivious transfer: Optimal protocols and their experimental implementations}.
\bjtitle{PRX Quantum}
\bvolume{4},
\bfpage{020320}
(\byear{2023})
\doiurl{10.1103/PRXQuantum.4.020320}
\end{barticle}
\endbibitem

\bibitem[\protect\citeauthoryear{Huang et~al.}{2012}]{NDSS:HuaEvaKat12}
\begin{bchapter}
\bauthor{\bsnm{Huang}, \binits{Y.}},
\bauthor{\bsnm{Evans}, \binits{D.}},
\bauthor{\bsnm{Katz}, \binits{J.}}:
\bctitle{Private set intersection: Are garbled circuits better than custom protocols?}
In: \bbtitle{{ISOC} Network and Distributed System Security Symposium -- {NDSS}~2012}.
\bpublisher{The Internet Society},
\blocation{San Diego, CA, USA}
(\byear{2012})
\end{bchapter}
\endbibitem

\bibitem[\protect\citeauthoryear{Dong et~al.}{2013}]{CCS:DonCheWen13}
\begin{bchapter}
\bauthor{\bsnm{Dong}, \binits{C.}},
\bauthor{\bsnm{Chen}, \binits{L.}},
\bauthor{\bsnm{Wen}, \binits{Z.}}:
\bctitle{When private set intersection meets big data: an efficient and scalable protocol}.
In: \beditor{\bsnm{Sadeghi}, \binits{A.-R.}},
\beditor{\bsnm{Gligor}, \binits{V.D.}},
\beditor{\bsnm{Yung}, \binits{M.}} (eds.)
\bbtitle{ACM CCS 2013: 20th Conference on Computer and Communications Security},
pp. \bfpage{789}--\blpage{800}.
\bpublisher{{ACM} Press},
\blocation{Berlin, Germany}
(\byear{2013}).
\doiurl{10.1145/2508859.2516701}
\end{bchapter}
\endbibitem

\bibitem[\protect\citeauthoryear{Lupo et~al.}{2023}]{PhysRevResearch.5.033163}
\begin{barticle}
\bauthor{\bsnm{Lupo}, \binits{C.}},
\bauthor{\bsnm{Peat}, \binits{J.T.}},
\bauthor{\bsnm{Andersson}, \binits{E.}},
\bauthor{\bsnm{Kok}, \binits{P.}}:
\batitle{Error-tolerant oblivious transfer in the noisy-storage model}.
\bjtitle{Phys. Rev. Res.}
\bvolume{5},
\bfpage{033163}
(\byear{2023})
\doiurl{10.1103/PhysRevResearch.5.033163}
\end{barticle}
\endbibitem

\bibitem[\protect\citeauthoryear{Brassard et~al.}{2000}]{brassard2000limitations}
\begin{barticle}
\bauthor{\bsnm{Brassard}, \binits{G.}},
\bauthor{\bsnm{L\"utkenhaus}, \binits{N.}},
\bauthor{\bsnm{Mor}, \binits{T.}},
\bauthor{\bsnm{Sanders}, \binits{B.C.}}:
\batitle{Limitations on practical quantum cryptography}.
\bjtitle{Phys. Rev. Lett.}
\bvolume{85},
\bfpage{1330}--\blpage{1333}
(\byear{2000})
\end{barticle}
\endbibitem

\bibitem[\protect\citeauthoryear{Hwang}{2003}]{hwang2003quantum}
\begin{barticle}
\bauthor{\bsnm{Hwang}, \binits{W.-Y.}}:
\batitle{Quantum key distribution with high loss: toward global secure communication}.
\bjtitle{Phys. Rev. Lett.}
\bvolume{91}(\bissue{5}),
\bfpage{057901}
(\byear{2003})
\end{barticle}
\endbibitem

\bibitem[\protect\citeauthoryear{Wang}{2005}]{wang2005beating}
\begin{barticle}
\bauthor{\bsnm{Wang}, \binits{X.-B.}}:
\batitle{Beating the photon-number-splitting attack in practical quantum cryptography}.
\bjtitle{Phys. Rev. Lett.}
\bvolume{94}(\bissue{23}),
\bfpage{230503}
(\byear{2005})
\end{barticle}
\endbibitem

\bibitem[\protect\citeauthoryear{Lo et~al.}{2005}]{lo2005decoy}
\begin{barticle}
\bauthor{\bsnm{Lo}, \binits{H.-K.}},
\bauthor{\bsnm{Ma}, \binits{X.}},
\bauthor{\bsnm{Chen}, \binits{K.}}:
\batitle{Decoy state quantum key distribution}.
\bjtitle{Phys. Rev. Lett.}
\bvolume{94}(\bissue{23}),
\bfpage{230504}
(\byear{2005})
\end{barticle}
\endbibitem

\bibitem[\protect\citeauthoryear{Kolesnikov et~al.}{2016}]{CCS:KKRT16}
\begin{bchapter}
\bauthor{\bsnm{Kolesnikov}, \binits{V.}},
\bauthor{\bsnm{Kumaresan}, \binits{R.}},
\bauthor{\bsnm{Rosulek}, \binits{M.}},
\bauthor{\bsnm{Trieu}, \binits{N.}}:
\bctitle{Efficient batched oblivious {PRF} with applications to private set intersection}.
In: \beditor{\bsnm{Weippl}, \binits{E.R.}},
\beditor{\bsnm{Katzenbeisser}, \binits{S.}},
\beditor{\bsnm{Kruegel}, \binits{C.}},
\beditor{\bsnm{Myers}, \binits{A.C.}},
\beditor{\bsnm{Halevi}, \binits{S.}} (eds.)
\bbtitle{ACM CCS 2016: 23rd Conference on Computer and Communications Security},
pp. \bfpage{818}--\blpage{829}.
\bpublisher{{ACM} Press},
\blocation{Vienna, Austria}
(\byear{2016}).
\doiurl{10.1145/2976749.2978381}
\end{bchapter}
\endbibitem

\bibitem[\protect\citeauthoryear{Haenni et~al.}{2017}]{haenni2017cast}
\begin{bchapter}
\bauthor{\bsnm{Haenni}, \binits{R.}},
\bauthor{\bsnm{Koenig}, \binits{R.E.}},
\bauthor{\bsnm{Dubuis}, \binits{E.}}:
\bctitle{Cast-as-intended verification in electronic elections based on oblivious transfer}.
In: \bbtitle{Electronic Voting: First International Joint Conference, E-Vote-ID 2016, Bregenz, Austria, October 18-21, 2016, Proceedings 1},
pp. \bfpage{73}--\blpage{91}
(\byear{2017}).
\bcomment{Springer}
\end{bchapter}
\endbibitem

\bibitem[\protect\citeauthoryear{Shor}{1994}]{FOCS:Shor94}
\begin{bchapter}
\bauthor{\bsnm{Shor}, \binits{P.W.}}:
\bctitle{Algorithms for quantum computation: Discrete logarithms and factoring}.
In: \bbtitle{35th Annual Symposium on Foundations of Computer Science},
pp. \bfpage{124}--\blpage{134}.
\bpublisher{{IEEE} Computer Society Press},
\blocation{Santa Fe, NM, USA}
(\byear{1994}).
\doiurl{10.1109/SFCS.1994.365700}
\end{bchapter}
\endbibitem

\bibitem[\protect\citeauthoryear{Bernstein and Lange}{2017}]{bernstein2017post}
\begin{barticle}
\bauthor{\bsnm{Bernstein}, \binits{D.J.}},
\bauthor{\bsnm{Lange}, \binits{T.}}:
\batitle{Post-quantum cryptography}.
\bjtitle{Nature}
\bvolume{549}(\bissue{7671}),
\bfpage{188}--\blpage{194}
(\byear{2017})
\end{barticle}
\endbibitem

\bibitem[\protect\citeauthoryear{Grilo et~al.}{2021}]{EC:GLSV21}
\begin{bchapter}
\bauthor{\bsnm{Grilo}, \binits{A.B.}},
\bauthor{\bsnm{Lin}, \binits{H.}},
\bauthor{\bsnm{Song}, \binits{F.}},
\bauthor{\bsnm{Vaikuntanathan}, \binits{V.}}:
\bctitle{Oblivious transfer is in {MiniQCrypt}}.
In: \beditor{\bsnm{Canteaut}, \binits{A.}},
\beditor{\bsnm{Standaert}, \binits{F.-X.}} (eds.)
\bbtitle{Advances in Cryptology -- {EUROCRYPT}~2021, Part~II}.
\bsertitle{Lecture Notes in Computer Science},
vol. \bseriesno{12697},
pp. \bfpage{531}--\blpage{561}.
\bpublisher{Springer},
\blocation{Zagreb, Croatia}
(\byear{2021}).
\doiurl{10.1007/978-3-030-77886-6_18}
\end{bchapter}
\endbibitem

\bibitem[\protect\citeauthoryear{Bose and Ray-Chaudhuri}{1960}]{bose1960class}
\begin{barticle}
\bauthor{\bsnm{Bose}, \binits{R.C.}},
\bauthor{\bsnm{Ray-Chaudhuri}, \binits{D.K.}}:
\batitle{On a class of error correcting binary group codes}.
\bjtitle{Information and control}
\bvolume{3}(\bissue{1}),
\bfpage{68}--\blpage{79}
(\byear{1960})
\end{barticle}
\endbibitem

\bibitem[\protect\citeauthoryear{Ishai et~al.}{2003}]{C:IKNP03}
\begin{bchapter}
\bauthor{\bsnm{Ishai}, \binits{Y.}},
\bauthor{\bsnm{Kilian}, \binits{J.}},
\bauthor{\bsnm{Nissim}, \binits{K.}},
\bauthor{\bsnm{Petrank}, \binits{E.}}:
\bctitle{Extending oblivious transfers efficiently}.
In: \beditor{\bsnm{Boneh}, \binits{D.}} (ed.)
\bbtitle{Advances in Cryptology -- {CRYPTO}~2003}.
\bsertitle{Lecture Notes in Computer Science},
vol. \bseriesno{2729},
pp. \bfpage{145}--\blpage{161}.
\bpublisher{Springer},
\blocation{Santa Barbara, CA, USA}
(\byear{2003}).
\doiurl{10.1007/978-3-540-45146-4_9}
\end{bchapter}
\endbibitem

\bibitem[\protect\citeauthoryear{Kolesnikov and Kumaresan}{2013}]{C:KolKum13}
\begin{bchapter}
\bauthor{\bsnm{Kolesnikov}, \binits{V.}},
\bauthor{\bsnm{Kumaresan}, \binits{R.}}:
\bctitle{Improved {OT} extension for transferring short secrets}.
In: \beditor{\bsnm{Canetti}, \binits{R.}},
\beditor{\bsnm{Garay}, \binits{J.A.}} (eds.)
\bbtitle{Advances in Cryptology -- {CRYPTO}~2013, Part~II}.
\bsertitle{Lecture Notes in Computer Science},
vol. \bseriesno{8043},
pp. \bfpage{54}--\blpage{70}.
\bpublisher{Springer},
\blocation{Santa Barbara, CA, USA}
(\byear{2013}).
\doiurl{10.1007/978-3-642-40084-1_4}
\end{bchapter}
\endbibitem

\bibitem[\protect\citeauthoryear{Keller et~al.}{2015}]{C:KelOrsSch15}
\begin{bchapter}
\bauthor{\bsnm{Keller}, \binits{M.}},
\bauthor{\bsnm{Orsini}, \binits{E.}},
\bauthor{\bsnm{Scholl}, \binits{P.}}:
\bctitle{Actively secure {OT} extension with optimal overhead}.
In: \beditor{\bsnm{Gennaro}, \binits{R.}},
\beditor{\bsnm{Robshaw}, \binits{M.J.B.}} (eds.)
\bbtitle{Advances in Cryptology -- {CRYPTO}~2015, Part~I}.
\bsertitle{Lecture Notes in Computer Science},
vol. \bseriesno{9215},
pp. \bfpage{724}--\blpage{741}.
\bpublisher{Springer},
\blocation{Santa Barbara, CA, USA}
(\byear{2015}).
\doiurl{10.1007/978-3-662-47989-6_35}
\end{bchapter}
\endbibitem

\bibitem[\protect\citeauthoryear{Beaver}{1992}]{C:Beaver91b}
\begin{bchapter}
\bauthor{\bsnm{Beaver}, \binits{D.}}:
\bctitle{Efficient multiparty protocols using circuit randomization}.
In: \beditor{\bsnm{Feigenbaum}, \binits{J.}} (ed.)
\bbtitle{Advances in Cryptology -- {CRYPTO}'91}.
\bsertitle{Lecture Notes in Computer Science},
vol. \bseriesno{576},
pp. \bfpage{420}--\blpage{432}.
\bpublisher{Springer},
\blocation{Santa Barbara, CA, USA}
(\byear{1992}).
\doiurl{10.1007/3-540-46766-1_34}
\end{bchapter}
\endbibitem

\end{thebibliography}
